\UseRawInputEncoding
\documentclass[lettersize,journal]{IEEEtran}
\usepackage{amsmath,amsfonts}
\usepackage{algorithmic}
\usepackage{algorithm}
\usepackage{array}
\usepackage[caption=false,font=normalsize,labelfont=sf,textfont=sf]{subfig}
\usepackage{textcomp}
\usepackage{stfloats}
\usepackage{url}
\usepackage{verbatim}
\usepackage{graphicx}
\usepackage{cite}
\newcommand{\RR}{\mathbb{R}}
\newcommand{\cF}{{\mathcal F}}
\newcommand{\cG}{{\mathcal G}}
\newcommand{\cH}{{\mathcal H}}

\begin{document}

\title{PR-DAD: Phase Retrieval Using Deep Auto-Decoders}

\author{Leon Gugel and Shai Dekel,School of mathematical sciences, Tel-Aviv university
}

\markboth{Journal of \LaTeX\ Class Files,~Vol.~xxx, No.~xxx, xxx~2022}%
{Shell \MakeLowercase{\textit{et al.}}: A Sample Article Using IEEEtran.cls for IEEE Journals}


\maketitle

\begin{abstract}
Phase retrieval is a well known ill-posed inverse problem where one tries to recover images given only the magnitude values of their Fourier transform as input. 
In recent years, new algorithms based on deep learning have been proposed, providing breakthrough results that surpass the results of the classical 
methods. In this work we provide a novel deep learning architecture PR-DAD (Phase Retrieval Using Deep Auto-Decoders), whose components are carefully designed based on mathematical modeling of the phase retrieval problem. The architecture provides experimental results that surpass all current results. 
\end{abstract}


\section{Introduction} \label{sec:introduction}

\subsection{The Phase Retrieval problem and classical methods}

The two-dimensional discrete Fourier transform $\cF(x)$ of an image $x\in\RR^{n\times n}$, can be represented by the magnitude 
\[
\omega(x):=|\cF(x)|\in\RR^{n\times n},
\]
and the phase
\[
\varphi(x):=\arg \cF(x)\in[-\pi,\pi]^{n\times n},
\]
where $\arg M$ denotes the argument of a complex matrix $M$ applied element-wise. The Fourier phase retrieval is a famous ill-posed inverse problem where 
the goal is to recover $x$, or equivalently the phase $\varphi(x)$, using only a input the magnitude $\omega(x)$. 
The difficulty of the problem stems from the fact that the phase contains most of the information of the image. 
This problem arises in many areas in engineering and science and has a rich history tracing back to 1952 \cite{S}. 
Important examples for Fourier phase retrieval naturally appear in many optical settings
since optical sensors, such as a charge-coupled device (CCD) and the human eye, are insensitive to phase information of the light wave. A typical
example is coherent diffraction imaging (CDI) which is used in a variety of imaging techniques (see \cite{BBE} and references therein). In CDI, an object
is illuminated with a coherent electro-magnetic wave and the far-field intensity diffraction pattern is measured. This pattern is proportional to the
object's Fourier transform and therefore the measured data is proportional to its Fourier magnitude. Phase retrieval also played a key role in the development
of the DNA double helix model \cite{GL}. Additional examples for applications in which Fourier phase retrieval appear are X-ray crystallography, speech recognition, blind channel
estimation, astronomy, computational biology, alignment and blind deconvolution (see \cite{BBE} and references therein).

The classical techniques for phase retrieval are iterative methods such as the alternating projection (see the survey \cite{BBE}). The general scheme of the 
alternating projection at each step $k$ is 
\begin{enumerate}
\item [(i)] compute the Fourier transform $\cF({x}_k)$ of the current estimated image $x_k$, 
\item [(ii)] keep its phase information $\varphi({x}_k)$, and replace the magnitude by the known ground truth magnitude $\omega({x}_k)=\omega({x})$,
\item [(iii)] compute the inverse Fourier to obtain a temporary estimate $\tilde{x}_{k+1}$,
\item [(iv)] impose certain known constraints, if needed, on $\tilde{x}_{k+1}$ (e.g. real non-negative pixel values), to obtain ${x}_{k+1}$. 
\end{enumerate} 

The PhaseCut method \cite{WdAM} is based on the following minimization formulation for the the input modulus $\omega$, unknown image $x=\{x_{j,k}\}$ with unknown phase $\varphi=\{\varphi_{j,k}\}$
\[
\min_{x,\varphi} \|\cF(x)-\omega\cdot\varphi\|^2,\quad \textrm{s.t } |\varphi_{j,k}|=1, \forall j,k.
\] 
There are several ways to relax this formulation and derive from it a minimization problem in the phase only, especially if $x$ is known to be real.

\subsection{The learning setup}

When we apply learning methods to an inverse problem such as phase retrieval, we need to clarify if we are attempting to solve the problem in the 
supervised, semi-supervised or un-supervised setting. First observe that one can easily compute the Fourier magnitude values for any ground truth image
and therefore such pairs can be used for supervised training.   
\begin{enumerate}
  \item [-] {\bf Supervised:} In this case we provide the trained model access to pairs of Fourier magnitude inputs and their corresponding ground
  truth images. Using these pairs, one can apply  a loss function such as Mean Square Error (MSE) between the predicted and ground truth images that will 
drive the minimization of a gradient descent method during the training of the model. 
\item [-] {\bf Semi-supervised:} In this setting only a partial subset of the Fourier magnitude inputs has corresponding ground truth images. 
This may happen in cases where we have acquired the Fourier magnitude of data through an imaging process, but we do not have knowledge about the 
ground truth image, except perhaps for the fact that it in a certain given image class with certain characteristics. This typically implies 
that to use the Fourier magnitude inputs which have no matching ground truth pixels during the training process, one needs to add additional 
loss mechanisms. One such loss function is the cycle loss which computes the Fourier magnitude of the images generated by the model
and then compares them with the input Fourier magnitudes. Another loss is the adversarial loss where a discriminator network is trained to provide a prediction 
if the image generated from the Fourier magnitude is plausible, i.e. if it belongs to the given class of images. 
\item [-] {\bf Un-supervised:} Here, we work with a dataset that has only Fourier magnitude inputs with no ground truth images at all. 
In this case we can only use loss functions such as the cycle loss to drive the training of our model. One can not use the adversarial loss since
there are no ground truth images that can be used as reference for the discriminator. However, if there exists some general prior knowledge on the structure of 
the given class of images, such as sparse Gaussian blobs, one can potentially transfer this knowledge into the form of a regularization loss function on the model's output images during training. 
\end{enumerate}   

In this work we assume that we are in the supervised or semi-supervised regimes, where we have a sufficient amount of training image samples from the given class. 
To achieve this, in some cases, we augment the given ground truth images by applying certain carefully designed transformations, thereby enriching the training set. 
One can also envision enriching the training dataset by creating synthetic data that faithfully represents the given class. 
For example, in the setting of x-ray crystallography one can generate synthetic virtual molecules from which one can compute pixel image slices.  
In this work, our fundamental assumption is that an inverse method based on learning can truly outperform the classical methods on a given class of images, 
if that class has sufficient structure and the learning algorithm can be trained to 'understand' that structure.

\subsection{Overview of Recent Deep Learning based methods}

We now review some recent work where deep learning methods are applied to the problem of phase retrieval. 

The DeepPhaseCut architecture \cite{CLJY} starts with a modified U-net generator $\cG_\Theta$ that takes as input the Fourier magnitude and predicts the
Fourier phase. We note in passing that applying a convolutional network, such as a U-net, on a frequency representation, is perhaps not optimal, since
there are typically no spatial correlations between `neighboring' Fourier coefficients or their respective magnitude. The predicted phase is then 
multiplied by the given magnitude to give a predicted Fourier transform. Then, an inverse Fourier transform is applied to obtain a predicted intermediate image.
The intermediate image is then fed into an enhancement network $\cH_\Psi$ to obtain the predicted image. The network is trained using several losses 
such as a cycle consistency loss, where the Fourier magnitude is extracted from the predicted image and compared to the input magnitude. 
The authors also trained discriminator classification networks that provide a score relating to the belief that an image is a ground truth image or 
an image produced by the generator network from Fourier magnitude data. This allows to use adversarial loss
during training. Lastly, although the DeepPhaseCut is somewhat equipped with a adversarial networks and a combination of adversarial and cycle loss function to deal with
the unsupervised phase retrieval problem, it also employs during the training process a supervised cycle loss component between ground truth images and the predicted images.    
 
In \cite{UOH}, similar concepts were used, namely, a generator was trained to take as input the Fourier magnitude and output a predicted image. The generator was trained 
with a linear combination of conditional and advesarial losses. Here as well, a discriminator was trained simultaneously to provide the advesarial loss.
The authors of \cite{UOH} note that a generator architecture based on fully connected layers
provides better empirical results than a convolutional architecture, which aligns with our understanding.     

In \cite{UHH} the authors propose to use a Cascaded Phase Retrieval (CPR) neural network (NN) architecture consisting of a sequence of sub-networks 
$G^{(1)},\dots,G^{(q)}$. Each subnetwork $G^{(i+1)}$ is fed as input the known magnitude $\omega(x)$ and $\hat{x}^{(i)}\in\RR^{n_i\times n_i}$, an estimate of the image
at some given (lower) resolution which is the output of the subnet $G^{(i)}$. The last subnet $G^{(q)}$ predicts the image $x$ at the full
resolution. The CPR network is trained with a loss function that incorporates all of 
the elements of the sequence of multiresolution approximations $\{\hat{x}^{(i)}\}$.

\section{The Auto-Encoder/Decoder Network} \label{sec:auto-encoder-decoder}

As already stated, in this work we assume that we have a sufficient amount of training image samples from the given class which allows us to first learn
some aspects of the structure of the class during a preprocessing stage. To be more specific, for each given class we design or train a representation
space in which the images of the class have a sparse representation. For piecewise constant images such as the MNIST dataset \cite{lecun1998gradient},
one can use the Haar wavelet orthonormal basis representation. For the fashion-MNIST dataset \cite{xiao2017fashion}, whose images also have some textured regions, we use the Haar wavelet packet basis representation (see Subsection \ref{subsec:fixed} below). For real-life image classes such the celebA \cite{liu2015deep}, we train carefully designed auto-encoder/decoder DL architectures
(see Subsection \ref{subsec:encoder-decoder} below). 
 
Once we find a good encoder-decoder pair that provides sparse representation for the image class, we extract the decoder part and plug it into our phase retrieval 
inference network. The central idea of our phase retrieval architecture is to use the prior knowledge about the encoded sparse structure of the class and ensure the network first maps the input Fourier magnitude to this space. Once this representation is obtained by the first part of the network, 
it undergoes enhancement and then is auto-decoded by the pre-trained decoder to provide the output approximate image. It is crucial to observe that it is 
the existence of the pre-trained decoder component that `forces' the network to transform the input Fourier magnitude to the desired encoded form.  
As we explained, there are two options to obtain an encoder-decoder pair that is adapted to the image class: using a  carefully selected fixed transform or training a neural network 
architecture:  

\subsection{Encoding using a fixed transform} \label{subsec:fixed} In some cases, one can simply 
select a certain transformation and its inverse as the encoder-decoder pair. This is especially useful in the semi-supervised or unsupervised settings where we 
do not have enough ground truth images to train an encoder-decoder network. There are three main properties that such a choice should satisfy:
\begin{enumerate}
  \item [(i)] Sparsity - The images from the given class should be sparse in the given transformed representation. Let $T(x)=\{\alpha_k(x)\}_k$
  be a transformation of an input image $x$ into a coefficient representation. Then a popular choice for a sparsity measure is requiring the $l_1$
norm $||T(x)||_1=\sum_k |\alpha_k(x)|$, to be minimal.  
\item [(ii)] Automatic differentiation of the inverse transform - The implementation of the inverse transform $T^{-1}$ needs to be plugged into a neural network,
such as the PR-DAD network, as a sub-network and undergo backpropogation during the training of the rest of the network.
\item [(iii)] Stability of the inverse transform - The inverse transform $T^{-1}$ should satisfy some stability condition, such as the frame condition 
\begin{equation} \label{frame-cond}
A||\alpha||_2^2 \le ||T^{-1}\alpha||_2^2 \le B||\alpha||_2^2, \quad\forall \alpha=\{\alpha_k\}_k,
\end{equation} 
where $0<A\le B <\infty$. This ensures stability of the backpropogation process during training. 
\end{enumerate}
Let us now provide some concrete examples for such transformations. An image class such as the MNIST dataset \cite{lecun1998gradient} of small grayscale hand-written
digits is a prototype example for a class of piecewise constant functions. It is well know that the Haar wavelet representation \cite{Dau} provides a sparse
representation for such images. Its simple implementation supports automatic differentiation and in this special case of an orthonormal transformation one may select $A=B=1$ in (\ref{frame-cond}). 

\begin{figure}[htbp] 
\centering
\includegraphics[width=3.5in]{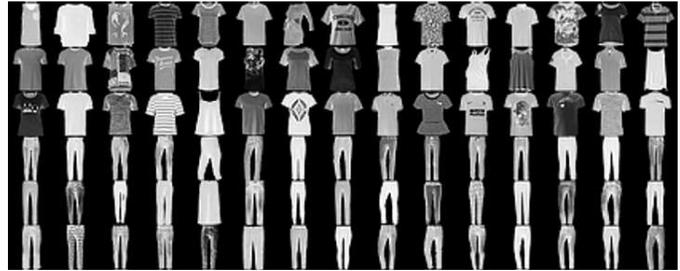}
\caption{Samples from the Fashion MNIST dataset}
\label{fig:fashionMNIST}
\end{figure}

\begin{figure}[htbp] 
\centerline{\includegraphics[width=3.5in]{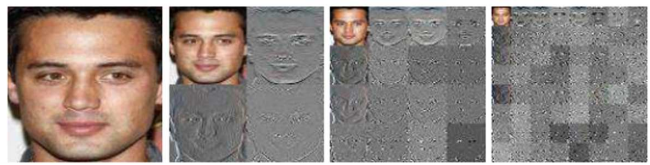}}
\caption{Haar Wavelet Packet transform (from \cite{HHST})}
\label{fig:HaarPacket}
\end{figure}

The fashion-MNIST dataset \cite{xiao2017fashion} consists of small grayscale images of clothes and accessories such as t-shirts, trousers, hand bags and shoes (some samples are 
depicted in Figure \ref{fig:fashionMNIST}). Some of the 
items, such as the t-shirts contain some texture components. An orthonormal transform that provides better sparsity for such data is the Haar wavelet packet transform 
depicted in Figure \ref{fig:HaarPacket} (see e.g. \cite{LF} for texture classification using the wavelet packet transform).
Compared to the more basic Haar transform, the packet transform further decomposes the subbands of the basic transform to time-frequency elements, which better capture
the local texture patches, i.e. allow a sparser representation. With the Haar packet transform a given image is decomposed into 4 subband blocks.
First we filter each row using the low-pass and high-pass filters
\[
\left(\frac{1}{\sqrt{2}},\frac{1}{\sqrt{2}}\right), \qquad \left(\frac{1}{\sqrt{2}},-\frac{1}{\sqrt{2}}\right),
\] 
and down-sample by a factor of 2. For an image of dimension $2^n\times 2^n$, this process gives a low-pass block and high-pass block, each of dimension
$2^{n-1}\times 2^n$. Next we filter each column in the same manner, thereby obtaining 4 blocks, each of dimension $2^{n-1}\times 2^{n-1}$. The 4
blocks are sometimes labeled by: LL,LH,HL and HH, where L=Low and H=high. In the packet transform these blocks are recursively filtered and subdivided,
where the last blocks that are decomposed are of dimension $2\times 2$.      

\subsection{Training an encode-decoder pair} \label{subsec:encoder-decoder} 
\begin{figure*}[htbp] 
\centerline{\includegraphics[width=6.5in]{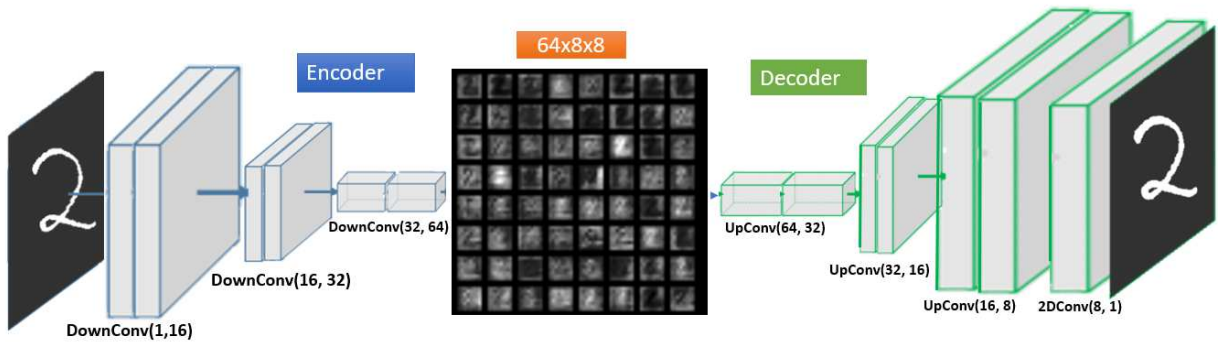}}
\caption{Encoder-decoder architecture: the case of low resolution dictionary}
\label{fig:encoder-decoder}
\end{figure*}

\begin{figure*}[htbp] 
\centerline{\includegraphics[width=5in]{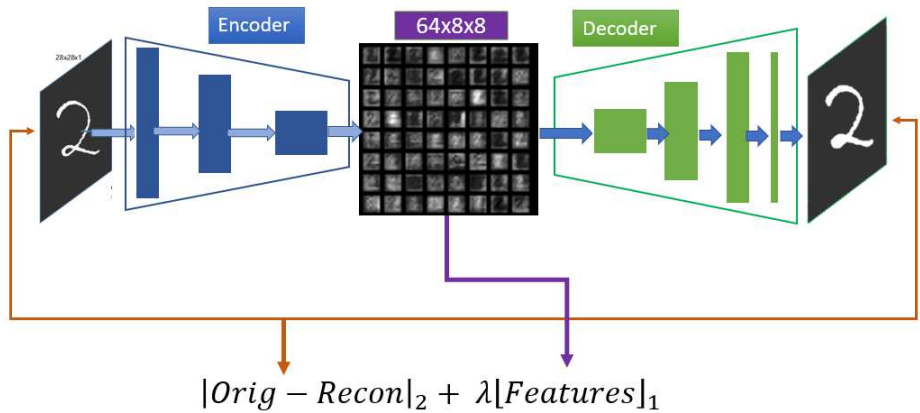}}
\caption{Encoder-decoder: training loss function}
\label{fig:encoder-decoder-loss}
\end{figure*}

A robust alternative to pre-selection of a fixed encoder-decoder pair, is to train such a pair, with the 
goal of learning a set of nonlinear projections of the images from the given class onto a sparse representation space. 
Here, we review one such useful architecture where the encoding space has a structure of over-redundant low-resolution
components. For a class of images of dimensions  $32\times 32$, the encoding space is composed of $64$ or $128$ feature maps, each of dimension $8\times 8$. 
In such a case, although the dimension of the encoding space can be larger than the image dimensions, the goal is that only a small portion of the encoding neurons have significant activations for a given image.   
In Figure \ref{fig:encoder-decoder} we see a depiction of the encoder-decoder architecture. The encoder part consists of 3 `DownConv' blocks. 
The `DownConv' block consists of 2 convolutional layers, 
each with convolutions of size $3\times 3$, stride$=1$, batch normalization and the ReLU or PReLU non-linear activations. Each `DownConv'
block concludes with an average pooling operation of $2\times 2$. The decoder subnet that recovers an image from the sparse representation has an almost
symmetric architecture. The first 3 blocks are upscaling `UpConv' blocks, where each block is composed of an upsampling bilinear interpolation operator,
followed by 2 convolution layers identical to the encoder convolution layers. The final decoder block consists of 1 convolutional layer whose output is 
the decoded image. Figure \ref{fig:encoder-decoder-loss} depicts the training loss function used for training the encoder-decoder architecture. 
It combines the $l_1$ sparsity regularization of the encoded representation space with the Mean Squared Error (MSE) loss between the 
input images to the encoder and the outputs of the decoder.

\section{The PR-DAD architecture}

\begin{figure*}[htbp] 
\centerline{\includegraphics[width=7in]{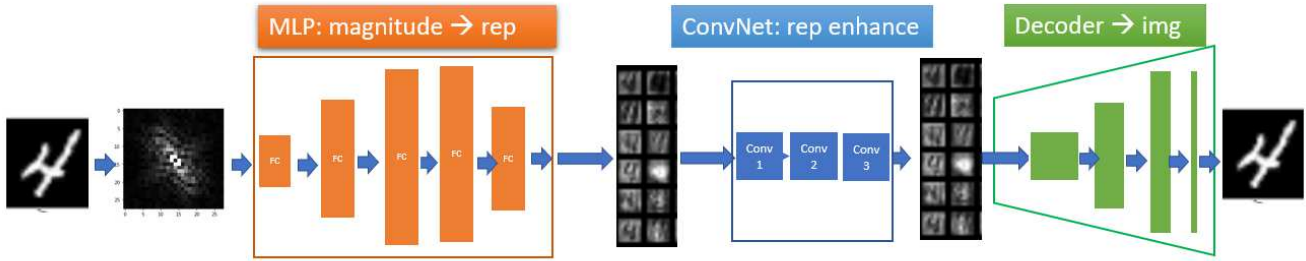}}
\caption{PR-DDL architecture}
\label{fig:architecture}
\end{figure*}

Figure \ref{fig:architecture} depicts the components of our PR-DAD architecture. It compromises of 3 subnets:
\begin{enumerate}
  \item [(i)] The Fourier magnitude to encoder representation subnet - this subnet is designed to take as input the Fourier magnitude, which is frequency related data and convert it to the pre-selected or pre-trained representation space. 
  \item [(ii)] The encoder representation enhancement subnet - the role of this subnet is to enhance the encoded representation before it is fed into the 
  decoder. 
  \item [(iii)] The decoder - this is the pre-selected or pre-trained decoder component that is plugged into the PR-DAD architecture. It is crucial 
  to understand that it is the decoder that drives the training process, in the sense that the two subnets leading to it need to provide the
decoder with a good approximation of the representation of the ground truth image, so as to minimize the training loss (see Subsection 
\ref{subsec:train-loss} for the training loss functions).   
\end{enumerate}

Observe that we do not attempt to reconstruct the actual phase at any stage of the network. Indeed, in our experiments, we found out that attempting to 
directly recover the phase so as to combine it with the known magnitude degrades the performance and is less stable then predicting a sparse representation from the magnitude, from which 
in turn one recovers the image using the decoder.

\subsection{The Fourier Magnitude to Encoder Representation Subnet} \label{subsec:MLP}
 
The goal of this subnet is to predict from the Fourier magnitude input, the encoder representation of the predicted image. 
In general, there is no immediate spatial correlation between neighboring Fourier coefficient values
in the frequency domain. Therefore, in contrast to some previous work, we prefer to process the input Fourier magnitude data using a relatively shallow 
Multi-Layer-Percepton (MLP) architecture, over a potentially deeper architecture of convolutional layers. In the MLP architecture, each layer contains a full  
affine transformation where any input value may contribute to any output value. In all our experiments we use an MLP consisting of 4 layers. It is important to point out that the nonlinear activation function we use is the Parametric ReLu (PReLu),
given by 

\[
\sigma_a(z):=
\begin{cases} z,  z>0, \\
az,  z\le 0,
\end{cases}
\]
where at each layer, the coefficient $a$ is a parameter of the network. The reason we do not use the ReLU activation function with the fixed 
parameter $a=0$, is that we need to ensure that the MLP subnet is consistent with the encoder representation
which may require negative values of representation neurons. For example, in the case where the encoder is selected to be the Haar transform, the 
representations are real wavelet coefficients with potentially negative values. 

The dimensions of the output of the last layer of the MLP are set to the dimensions of the auto-encoder
features.  For example, if we use a set of $128$ auto-encoder feature maps, each of dimension $8\times 8$ pixels, then the output of the last
MLP layer is then of dimension $128 \times 8 \times 8$.

It is interesting to note that initially we tried to use a fixed inverse Fourier or inverse Discrete Cosine Transform (DCT) component that will take in
the output of the MLP subnet, apply to it the inverse transform and then pass forward the data to the encoding representation layer. Such a component is 
used for example in the DeepPhaseCut algorithm \cite{CLJY} on output of a phase generator subnet after it is combined with the input magnitude. 
Our initial idea was that the MLP subnet will learn the Fourier transform of the encoding representation to which one should apply the inverse Fourier transform. 
However, in our architecture, since any linear transformation such as the inverse Fourier can be combined into the final affine transformation followed by a PReLU activation taking 
place in a MLP subnet, we found through ablation experimentation that the inverse Fourier component is not needed in our architecture, as it is in some
sense already `realized' by the MLP subnet.      

\subsection{The Encoding Representation Enhancement Subnet}

This is an optional subnet of the PR-DAD architecture (see Figure \ref{fig:architecture}). It assumes that the previous MLP subnet succeeded 
to convert the Fourier magnitude data to the encoding representation and its goal is to provide some capacity for the purpose of enhancing the representation,
before it is passed to the decoder. 

In the case where the encoding representation space was created by a convolutional encoder, the enhancement subnet is also convolutional. Let us assume 
that the representation space is composed of $N$ feature maps, each of dimension $n\times n$ (e.g. $N=128$, $n=8$). Then, the input and 
output of each layer of the enhancement network are $N$ channels/feature maps of dimension $n\times n$. We apply filters of $X-Y$ dimension 
$3\times 3$, which implies each filter is of dimension $N\times 3 \times 3$. This determines that each feature map is enhanced using
information from all other feature maps. Our implementation in the experimental results below uses 3 convolutional blocks, each consisting of 
2 convolutional layers, each with batch normalization and PReLU activation.   

\subsection{The  Decoder Subnet}

This subnet is initially a fixed component of the network, whose architecture is the decoder part of the auto-encoder/decoder network that was 
pre-selected or trained during the 
preprocessing stage described in Section \ref{sec:auto-encoder-decoder}. In the case where the decoder is trained, this subnet can use the fixed weights that were computed during the 
encoder-decoder training process. It is also possible to allow this subnet to participate in the training of the full PR-DAD network during the last few epochs, following a `transfer learning' paradigm.
Through experimentation, we found that this could slightly improve the performance in some cases.     

\subsection{The PR-DAD Training Loss Functions} \label{subsec:train-loss}

\begin{figure*}[htbp] 
\centerline{\includegraphics[width=7in]{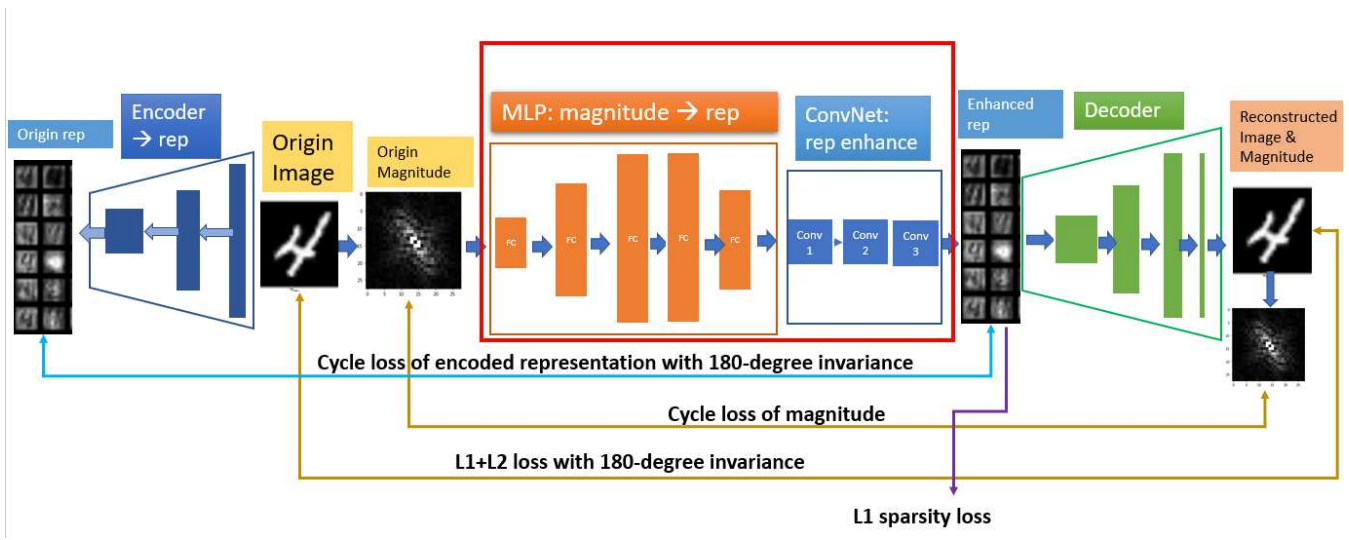}}
\caption{Loss functions used during training}
\label{fig:loss-functions}
\end{figure*}

Figure \ref{fig:loss-functions} depicts all of the loss function components used for the training of the PR-DAD architecture, where the actual loss function 
is a weighted sum of all of them. The following loss functions are applied for each training batch $B=\{x_i\}_{i\in B}$

\begin{enumerate}
  \item [(i)] MSE loss of predicted images $\{\hat{x}_i\}_{i\in B}$, invariant to rotation by $\pi$
\[
L_{MSE} = \frac{1}{\#B} \sum_{i\in B} \min(||x_i - \hat{x}_i ||_2^2,||x_i - rotate_\pi(\hat{x}_i) ||_2^2).
\]
  \item [(ii)] Cycle loss with predicted magnitude 
\[
L_{mag} = \frac{1}{\#B} \sum_{i\in B} ||\omega(x_i) - \omega(\hat{x}_i) ||_2^2.
\]
\item [(iii)] Sparsity of predicted encoding representations $\{\hat{T}_i\}_{i\in B}$
\[
L_{sparse} = \frac{1}{\#B} \sum_{i\in B} ||\hat{T}_i ||_1.
\]
\item [(iv)] MSE loss of predicted encoding representations $\{\hat{T}_i\}_{i\in B}$, invariant to rotation by $\pi$
\[
L_{encode} = \frac{1}{\#B} \sum_{i\in B} \min(||T_i-\hat{T}_i||_2^2,||T_i-rotate_\pi(\hat{T}_i)||_2^2).
\]  
Observe that the rotation of the representation space is an operation depending on the encoding method. For example, in the case of a 
representation by $N$ `low resolution' elements of dimensions $n\times n$, the rotation operation is applied separately on each element.   
\end{enumerate} 
Then, the training loss is a weighted sum of all of the above losses 
\[
L = \lambda_{MSE} L_{MSE} + \lambda_{mag} L_{mag} + \lambda_{sparse} L_{sparse} + \lambda_{encode} L_{encode}
\]

\section{Experimental Results} \label{sec:results}

\subsection{Overview of Datasets}

For the experimental evaluation we used four ``MNIST'' datasets, each consisting of grayscale images of dimension $28\times 28$: 
\begin{enumerate}
  \item [(i)] MNIST \cite{lecun1998gradient} - 70,000 images of hand written digits from 10 classes $1-10$,
  \item [(ii)] EMNIST \cite{cohen2017emnist} - The balanced version of 131,600 images containing hand written letters and digits from 47 classes, 
  \item [(iii)] Fashion-MNIST \cite{xiao2017fashion} - 70,000 images of clothing from 10 classes such as: T-shirt, Trouser, Pullover, etc., 
  \item [(iv)] KMNIST \cite{clanuwat2018deep} - 70,000 images of 10 types of handwritten Japanese characters.
\end{enumerate} 
The 5th dataset we used is the more challenging CelebA dataset \cite{liu2015deep}, which consists of
200,000 images of human faces (20 images of 10,000 different individuals in diverse poses and setting). The datasets underwent certain pre-processing for 
two purposes: We applied exactly the same cropping and resizing as in previous work, so we can compare the results (see tables below). We enriched the 
training sets using certain transformations. Table \ref{tbl:data-prep} summarizes all of the transformations for each given dataset. 

\begin{table*}[htbp] 
\begin{center}
\caption{Dataset pre-processing and augmentation.}
\begin{tabular}{|p{100pt}|l|l|l|l|p{50pt}|}
\hline
\textbf{Transform} &
\textbf{MNIST} &
\textbf{EMNSIT} &
\textbf{KMNIST} &
\textbf{Fashion Mnist} &
\textbf{CelebA}\\ 
\hline \hline
Resizing &
$32\times32$ &
$32\times32$ &
$32\times32$ &
$32\times32$ &
$64\times64$ \\
\hline

Center Cropping &
No &
No &
No &
No &
Yes \\
\hline

Normalization $(\mu, \sigma)$ &
$(0.1307, 0.3081)$ &
$(0.1307, 0.3081)$ &
$(0.1307, 0.3081)$   &
$(0.1307, 0.3081)$ &
$(0.5, 0.5)$ \\
\hline

Fourier Magnitude zero padding &
0.5 &
0.5 &
0.5 &
0.25 &
No \\
\hline

Bernoulli probability $\rho$ &
0.25 &
0.25 &
0.50 &
0.25 &
0.50\\
\hline

Random Horizontal Flipping with probability $\rho$ &
No &
No &
No &
No &
Yes \\
\hline

Random Free Rotation with probability $\rho$ in range $(\theta, -\theta)$ &
No &
No &
No &
$(1.0, 2.5)$ &
No \\
\hline

Random Free Translation with probability $\rho$ in range $\tau_1, \tau_2$ &
$(0.025, 0.025)$ &
$(0.025, 0.025)$ &
$(0.025, 0.025)$ &
$(0.0125, 0.025)$ &
$(0.025, 0.025)$\\
\hline

Random Free Scaling with probability $\rho$ in range $(r_1, r_2)$ &
$(0.9, 1.2)$ &
$(0.9, 1.2)$ &
$(0.9, 1.2)$&
$(0.95, 1.1)$ &
$(0.9, 1.2)$ \\
\hline

Random Gaussian Blur, prob $\rho$, kernel $k$ with $\sigma$ &
No &
No &
0.0 &
No &
$(0.5, 1.5)$\\
\hline

Random Gamma Correction, prob $\rho$ in range $(\gamma_1, \gamma_2)$ &
No &
No &
0.0 &
No &
$(0.85, 1.125)$\\
\hline

\end{tabular}
\label{tbl:data-prep}
\end{center}
\end{table*}

\subsection{Results in the Supervised Setting}

The implementation of the PR-DAD algorithm can be found on the GitHub \cite{Gu}. For the implementation of the Haar wavelet packet as an encoder-decoder option, 
we used some parts of the code from \cite{HHST}. The training of both encoder-decoder and the PR-DAD architectures were done using a V100 Tesla GPU. 
For the training we used batches of 32-64. We applied the Adam
stochastic gradient decent algorithm using the loss functions detailed in Subsection \ref{subsec:train-loss} 

The metrics used for evaluation are on par with the previous work. Given two images $x,\hat{x}$ of size $N$ we have: 
\begin{enumerate}
  \item [(i)] MSE loss - Mean Squared Error $\frac{1}{N}\sum_{i,j} (x_{i,j} - \hat{x}_{i,j})^2$, lower is better.
  \item [(ii)] MAE loss - Mean Average Error $\frac{1}{N}\sum_{i,j} |x_{i,j} - \hat{x}_{i,j}|$, lower is  better.
  \item [(iii)] SSIM - Structural Similarity, higher is better.  
  \item [(iv)] PSNR - Peak to Signal Noise Ratio $10\log_{10}\left(\frac{255^2}{MSE}\right)$, higher is better.
\end{enumerate}

\begin{figure*}[htbp] 
\centerline{\includegraphics[width=6in]{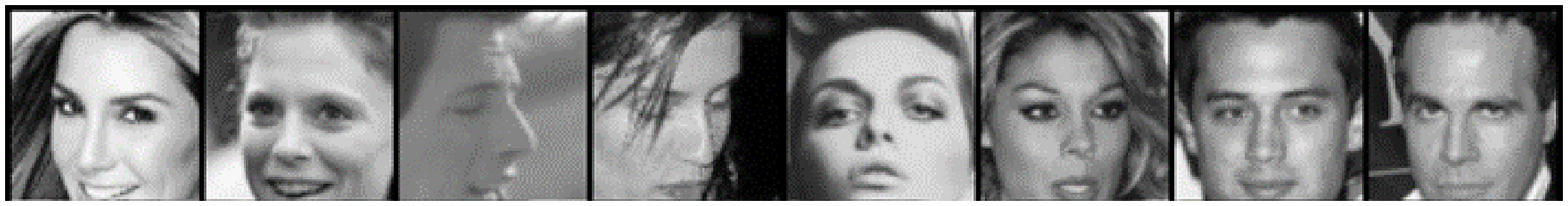}}
\centerline{\includegraphics[width=6in]{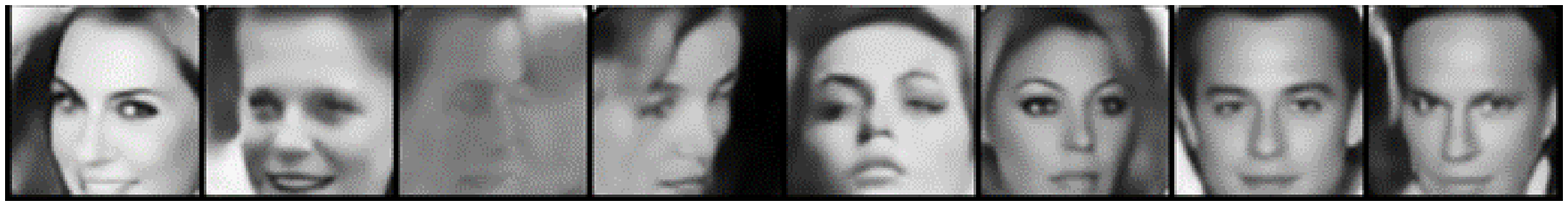}}
\caption{Top -  cropped CelebA original images, bottom - cropped celebA recovered images}
\label{fig:celebA-examples}
\end{figure*}

In Tables \ref{table-MNIST}-\ref{table-CELEBA} we see the test results on the MNIST, EMNIST, KMNIST, Fashion-MNIST and CelebA datasets.
In Figure \ref{fig:celebA-examples} we see some samples of pairs of original and recovered cropped celebA images. It is very evident that on the more 
challenging dataset such as the celebA dataset, the PR-DAD demonstrates superior performance. 

\begin{table}[htbp]
\begin{center}
\caption{Quantitative comparison on the MNIST dataset}
\begin{tabular}{|p{80pt}|l|l|l|p{30pt}|}
\hline
\textbf{Model} &
\textbf{MSE} &
\textbf{MAE} &
\textbf{SSIM}&
\textbf{PSNR} \\ 
\hline \hline
PRCGAN \cite{UOH} &
0.0168&
0.0399& 
0.8449&
- \\
\hline
CPR \cite{UHH} &
0.0123&
{\bf 0.037}& 
0.8756&
- \\
\hline
{\bf PR-DAD Haar Packet} &
0.0106&
0.0381& 
{\bf 0.8815}&
39.4861\\
\hline
{\bf PR-DAD auto encoder-decoder} &
{\bf 0.0100}&
0.0398& 
0.8799&
{\bf 40.0208}\\
\hline

\hline
\end{tabular}
\label{table-MNIST}
\end{center}
\end{table}

\begin{table}[htbp]
\begin{center}
\caption{Quantitative comparison on the EMNIST dataset}
\begin{tabular}{|p{80pt}|l|l|l|p{30pt}|}
\hline
\textbf{Model} &
\textbf{MSE} &
\textbf{MAE} &
\textbf{SSIM}&
\textbf{PSNR} \\ 
\hline \hline
PRCGAN \cite{UOH} &
0.0239&
0.0601& 
0.8082&
- \\
\hline
CPR \cite{UHH} &
0.0144&
0.0501& 
0.8700&
- \\
\hline
{\bf PR-DAD Haar Packet} &
0.0119&
0.0475& 
0.8710&
38.4744\\
\hline
{\bf PR-DAD auto encoder-decoder} &
\textbf{0.0108}&
\textbf{ 0.0422}& 
\textbf{ 0.8879}&
\textbf{39.2972}\\
\hline

\hline
\end{tabular}

\label{table-EMNIST}
\end{center}
\end{table}

\begin{table}[htbp]
\begin{center}
\caption{Quantitative comparison on the KMNIST dataset}
\begin{tabular}{|p{80pt}|l|l|l|p{30pt}|}
\hline
\textbf{Model} &
\textbf{MSE} &
\textbf{MAE} &
\textbf{SSIM}&
\textbf{PSNR} \\ 
\hline \hline
PRCGAN \cite{UOH} &
0.0651&
0.1166& 
0.5711&
- \\
\hline
CPR \cite{UHH} &
0.0433&
0.1034& 
0.6624&
- \\
\hline
{\bf PR-DAD Haar Packet} &
0.0383&
0.1027& 
0.6365&
28.3249\\
\hline
{\bf PR-DAD auto encoder-decoder} &
\textbf{0.0380}&
\textbf{0.0957}& 
\textbf{0.6605}&
\textbf{28.4031}\\
\hline

\hline
\end{tabular}

\label{table-KMNIST}
\end{center}
\end{table}

\begin{table}[htbp]
\begin{center}
\caption{Quantitative comparison on the Fashion-MNIST dataset}
\begin{tabular}{|p{80pt}|l|l|l|p{30pt}|}
\hline
\textbf{Model} &
\textbf{MSE} &
\textbf{MAE} &
\textbf{SSIM}&
\textbf{PSNR} \\ 
\hline \hline
PRCGAN \cite{UOH} &
0.0151&
0.0572& 
0.7749&
- \\
\hline
CPR \cite{UHH} &
0.0113&
0.0497& 
0.8092&
- \\
\hline
{\bf PR-DAD Haar Packet} &
{\bf 0.0078}&
0.0471& 
0.8186&
{\bf 42.1862}\\
\hline
{\bf PR-DAD auto encoder-decoder} &
0.0081&
{\bf 0.0442}& 
{\bf 0.8242}&
41.811\\
\hline

\hline
\end{tabular}

\label{table-FASHIONMNIST}
\end{center}
\end{table}

\begin{table}[htbp]
\begin{center}
\caption{Quantitative comparison on the cropped CelebA $64\times 64$ dataset}
\begin{tabular}{|p{77pt}|l|l|l|p{30pt}|}
\hline
\textbf{Model} &
\textbf{MSE} &
\textbf{MAE} &
\textbf{SSIM} &
\textbf{PSNR} \\ 
\hline \hline
PRCGAN \cite{UOH} &
0.0138&
0.0804& 
0.6779&
n/a\\
\hline

HIO \cite{Fi}&
n/a&
n/a& 
0.472&
19.573\\
\hline

PhaseCut \cite{WdAM} &
n/a&
n/a& 
0.7600&
25.3600\\
\hline

On-RED \cite{WSLK} &
n/a&
n/a& 
0.4940&
19.7960\\
\hline

PrDeep \cite{MS} &
n/a&
n/a& 
0.7380&
26.0579\\
\hline

DeepPhaseCut \cite{CLJY}&
n/a&
n/a& 
0.8540&
27.1190\\
\hline

{\bf PR-DAD} &
{\bf 0.0025}&
{\bf 0.0340}& 
{\bf 0.8815}&
{\bf 51.9661}\\
\hline

\end{tabular}
\label{table-CELEBA}
\end{center}
\end{table}

\section{Conclusions}

In this paper we presented a deep learning approach to the phase retrieval problem. The PR-DAD algorithm uses an encoder-decoder transform or network that provides a sparse representation of images from a given class. This facilitates solving the phase retrieval problem by an architecture that predicts from the Fourier magnitude data the image, without trying to predict the actual phase.
We showed that our solution provides experimental results that are highly competitive. 

In the future we plan to expand the capabilities of the algorithm and apply it on microscopy and crystallography datasets. Such datasets will potentially require different encoder-decoder 
architectures. We also plan to try and develop capabilities in a semi-supervised setting, where there exists only a small amount of ground truth images. This makes training an 
encoder-decoder architecture more difficult.

\newpage

\section{Biography Section}

\begin{IEEEbiographynophoto}{Shai Dekel}
Shai is a visiting associate professor at the school of mathematical sciences, Tel-Aviv university.  
\end{IEEEbiographynophoto}

\begin{IEEEbiographynophoto}{Leon Gugel}
Leon is a senior deep learning expert at D-ID  
\end{IEEEbiographynophoto}

\vfill


\begin{thebibliography}{1}
\bibliographystyle{IEEEtran}

\bibitem{BBE}
 T. Bendory, R. Beinert and Y. Eldar, Fourier Phase Retrieval: Uniqueness and Algorithms, In: Compressed Sensing and its Applications (2017), 55-91.

\bibitem{clanuwat2018deep} 
T. Clanuwat, M. Bober-Irizar, A. Kitamoto, A. Lamb, K. Yamamoto and D. Ha, Deep learning for classical Japanese literature, arXiv:1812.01718 (2018).

\bibitem{cohen2017emnist} 
G. Cohen, S. Afsher, J. Tapson and A. Schaik, EMNIST: Extending MNIST to handwritten letters, IEEE international joint conference on neural networks 2017.

\bibitem{CLJY}
 E. Cha, C. Lee, M. Jang and J Ye, DeepPhaseCut: deep relaxation in phase for unsupervised Fourier phase retrieval, IEEE Transactions on Pattern Analysis and Machine Intelligence, to appear.

\bibitem{Dau}
I. Daubechies, Ten lectures on wavelets, SIAM 1992.

\bibitem{Fi}
J. R. Fienup, Phase retrieval algorithms: a comparison, Applied optics 21 (1982), 2758-2769.

\bibitem{GL}
 L. Garwin and T. Lincoln, A century of nature: twenty-one discoveries that changed science and the world, University of Chicago
Press, 2010.

\bibitem{Gu} L. Gugel, PR-DAD code, https://github.com/gugas81/pr-dad.

\bibitem{HHST} H. Huang, R. He, Z. Sun, T. Tan, Wavelet-SRNet: A Wavelet-Based CNN for Multi-Scale Face Super Resolution, Proc. IEEE international 
Conference on Computer Vision (ICCV), 2017, 1689-1697.

\bibitem {LF} A. Laine and J. Fan, Texture classification by wavelet packet signatures, IEEE Transactions on pattern analysis and machine intelligence 15 (1993), 1186-1191.

\bibitem{lecun1998gradient}
 Y. LeCun, L. Bottou, Y. Bengio and P. Haffner, Gradient-based learning applied to document recognition, Proceedings of the IEEE 86 (1998), 2278-2324.

\bibitem{liu2015deep}
Z. Liu, P. Luo, X. Wang and X. Tang, Deep learning face attributes in the wild, Proceedings of the IEEE international conference on computer vision (2015),
3730-3738.

\bibitem{MS} C. Metzler, P. Schniter, A. Veeraraghavan and R. Baraniuk, prDeep: Robust phase retrieval with a flexible deep network, ICML 2018, 3501-3510.

\bibitem{S}
 D. Sayre, Some implications of a theorem due to Shannon. Acta Crystallographica, 5 (1952), 843-843.

\bibitem{UHH}
 T. Uelwer, T. Hoffmann and S. Harmeling, Non-iterative phase retrieval with cascaded neural networks, ICANN 2021.

\bibitem{UOH}
 T. Uelwer, A. Oberstraß and S. Harmeling, Phase retrieval using conditional generative adversarial networks, ICPR 2021.

\bibitem{WdAM}
 I. Waldspurger, A. d'Aspremont and S. Mallat, Phase recovery, maxcut and complex semidefinite programming, Mathematical Programming 149 (2015), 47-81.

\bibitem {WSLK} Z. Wu, Y. Sun, J. Liu, and U. Kamilov, Online regularization by denoising with applications to phase retrieval, Proc. IEEE International Conference
 on Computer Vision Workshops 2019,
pp. 0-0.

\bibitem{xiao2017fashion} 
H. Xiao,K. Rasul and R. Vollgraf, Fashion-mnist: a novel image dataset for benchmarking machine learning algorithms, arXiv:1708.07747 (2017).


\end{thebibliography}
\end{document}